\documentstyle[prb,eqsecnum,aps,epsf]{revtex}
\begin{document}
\draft
\setlength{\textfloatsep}{10pt plus 2 pt minus 2 pt}
\twocolumn[\hsize\textwidth\columnwidth\hsize\csname@twocolumnfalse%
\endcsname

\title{Density matrix renormalization group study of dimerization
of the Pariser-Parr-Pople model of polyacetilene.}
\author{G.L. Bendazzoli, S. Evangelisti}
\address{
  Dipartimento di Chimica Fisica e Inorganica, Universit\`a di Bologna\\
  viale Risorgimento 4, 40136 Bologna - Italy\\
}
\author{G. Fano, F. Ortolani and L. Ziosi}
\address{
  Dipartimento di Fisica, Universit\`a di Bologna\\
  Via Irnerio 46 , 40126 Bologna -  Italy\\
  fano@bologna.infn.it\\
}

\date{\today}
\maketitle

\begin{abstract}
We apply the DMRG method to the Pariser-Parr-Pople hamiltonian
and investigate the onset of dimerization. We deduce  the parameters of
the hopping term and the contribution of the $\sigma$ bonds
 from ab initio calculations on ethylene. Denoting by $R_{ij}$ the
$C-C$ distances, we perform a variational optimization of the
dimerization $\delta= (R_{i,i+1} - R_{i-1,i})/2$
 and of the average bond length
$R_0$ for chains up to $N=50$ sites. The critical value of $N$  at
which the transition occurs is found to be between $N=14$ and $N=18$
 for the present model.  The asymptotic values for large $N$
for $R_0$ and $\delta$ are  given by $1.408(3) \AA $ and $0.036(0) \AA$.
\bigskip
\end{abstract}

]

\bigskip
\section{Introduction}
The quantum chemical treatment of large polymeric systems is extremely
demanding from the computational point of view.  When the
electron correlation is taken into account, traditional correlated
quantum chemical methods grow too rapidly 
with the number of electrons ($n^5$ or worse)
to be considered as practical tools to study large systems.
Density Functional Theory (DFT) is a powerful tool to cope with this 
kind of problems,
and a number of efficient codes are available nowadays.
However, 
the local density approximation (even with gradient corrections)
can only treat the local part of the electron-electron correlations
(within an atom or  bond) and does not treat the interatomic
correlations with sufficient accuracy.
Moreover, no information on the wave 
function of the system is provided by this method.
In this paper
we focus our attention on a different method, i.e. 
the Density Matrix Renormalization Group
(DMRG) formalism, proposed by White\cite{white1}.
This approach in principle can also be developed to
an efficient tool to compute correlated energies and {\it wavefunctions} 
of the ground or
excited states of large systems.
DMRG is an approximation method that uses a density matrix to
identify the most relevant states of a full configuration 
interaction (FCI) expansion.
The localized orbital set is divided into two adjacent parts called ''system''
and ''environment''.
The DMRG method allows one  to determine in
a precise mathematical way the most probable states of
the ''system''  in the presence of the ''environment''. These states
are retained, and the others are dropped. The calculation starts with
two small fragments, whose size is increased 
in the course of the calculation, until the
whole system reaches the wanted dimension. We refer the reader
to Refs. [\onlinecite{white1,white2}] and [\onlinecite{fano}]
for a detailed description of the method. \par
DMRG has been applied to a wide class of systems, including
strongly-correlated electrons, 
antiferromagnetic Heisenberg chains\cite{white3},
phonons\cite{phonons,caron}, defined on one and two dimensional lattices
\cite{white_2dtj}.
Originally DMRG was designed to treat nearest-neighbor interactions
in one dimension, and in these conditions the method
is very efficient in reducing the computational effort. 

The Pariser-Parr-Pople (PPP) hamiltonian of conjugated polyenes \cite{PPPH} 
was recently studied with DMRG\cite{yaron,fano}.
In Ref. [\onlinecite{yaron}] the dimerized problem was considered with 
a symmetrized version of DMRG\cite{ramasesha}
that allows  the selection of excited states of given symmetry.
In Ref. [\onlinecite{fano}] the ground state at constant bond length 
was examined, and the results were in full agreement with FCI 
calculations, in all cases where these latter
were available\cite{stef1,stef2}.
The PPP hamiltonian has a long history
since in spite of its simplicity  it provides at least a  qualitative
explanation of the essential physics  of polyacetilene \cite{Chien,Soos}.  
We decided to consider the dimerization of the Pariser-Parr-Pople model
of conjugated polyenes. 
Systematic studies of the bond alternation in conjugated
polyenes were carried out by Paldus and coworkers \cite{Paldus,Paldus1}.
These authors used various kinds of approximations (self-consistent field,
alternant molecular orbitals, coupled cluster, valence bond) 
since the FCI treatment  ceases to be  applicable before the onset of the 
bond alternation. It seems therefore interesting to compare those results
with DMRG calculations on the same hamiltonian. 
Recently a Density Functional (DFT) computation on large
annulenes has been performed \cite{Kertesz}, so a comparison of our 
DMRG results obtained with a model hamiltonian with a realistic  all-electron 
calculation can give useful information on the properties of the PPP model.
DMRG has already been applied to the study of 
dimerization\cite{lepetit,solitons,peierls-hubbard,excitons}
of polyacetilene chains using various model hamiltonians.
In Ref. [\onlinecite{lepetit}], the authors considered
a generalization of the  Hubbard model to describe the
valence $\pi$ electrons and approximated the $\sigma$ contributions
by a sum of pair interactions. Their generalization of the Hubbard model
consisted of allowing for a distance-dependent hopping term,
while the Coulomb on-site repulsion was taken to be independent
of the distance between $C-C$ atoms. They determined the parameters of
the model from accurate {\it ab initio}  CI calculations on the ground state
and the first excited state of the ethylene molecule, available as a function
of the $C-C$ distance\cite{said}. The dimerization $\delta$ and the
average bond-length  $R$ were determined minimizing the energy. \par

The aim of this paper is to exploit the same method for the
determination of the parameters as a function of the $C-C$ distance,
and to improve the treatment of the correlation by substituting the
crude Hubbard approximation with the more accurate
PPP treatment of the Coulomb repulsion. For this purpose
a generalization of the PPP hamiltonian is used where both the
hopping integral and the potential depend on the  geometry of
the system. In this way, the values of the average $C-C$ distance
and of the bond-alternation can be determined
by a variational calculation as a function of $N$.
Furthermore different geometries can be easily considered,
which was not the case for the Hubbard model. \par

\section{Definition of  the model}

To describe dimerization in trans-polyacetilene chains
we introduce the following distance  dependent hamiltonian:
\begin{equation}
        H = H_\sigma + H_\pi
\end{equation}
The $\sigma$ contribution is approximated by a sum over all  bonds
and depends only on the bond lengths:
\begin{equation}
        H_\sigma  = \sum_{<i,j>} E_\sigma(R_{ij})
\end{equation}
($R_{ij}$ is the distance between two carbon atoms located
at sites $i$ and $j$ and $<ij>$ denotes summation restricted to
nearest neighbors).
The $\pi$ contribution is represented by the Pariser-Parr-Pople
hamiltonian, with distance dependent hopping:

\begin{figure}
\centerline{I}
\centerline{
\epsfbox{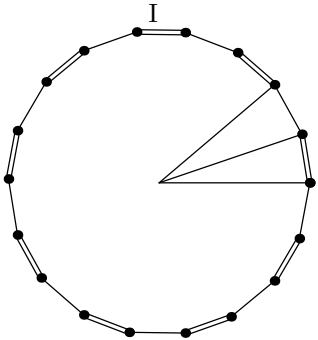}
}
\caption{Structure I: all C atoms are on a ring.
\label{fig:circumference}}
\end{figure}

\begin{equation}
        H_\pi    = \sum_{ < i j >} \beta_{i,j} \hat{E}_{i j}
          + {1\over 2} \sum_{i,j=0}^{N-1} \gamma_{i j}
           \left(\hat {n}_i - 1 \right)
           \left( \hat {n}_j - 1 \right)
\label{PPPham}
\end{equation}

Here $\hat{E}_{i j}$  are the generators of the unitary
group summed over spin, and $\hat {n}_i = \hat {E}_{i i}$
is the occupation number of the site $i$; $\beta_{i j}$,
$\gamma_{i j}$ are parameters of the model.
For the Coulomb repulsion we use the
Mataga-Nishimoto prescription  \cite{mataga}:
\begin{equation}
        \gamma_{i j} = {1 \over  \gamma_0^{-1} + R_{i j}}
        \qquad\hbox{(a. u.)}
\end{equation}
where  $\gamma_0 = 10.84$ eV.

Following Ref. [\onlinecite{lepetit}] we determine
$\beta_{i j}$ and  $H_\sigma$
by imposing that  the singlet triplet splitting 
calculated {\it ab initio} for
the ethylene molecule is correctly reproduced.
In  Ref. [\onlinecite{said}] the singlet-triplet splitting
and the sigma energy are given as
polynomial expansions in the bond length:
\begin{eqnarray}
g(R) \equiv & (\, ^3E(R) -\, ^1E(R) ) / 2 &=  \sum_{l=1}^5 a_l R^l \\
         &  E_\sigma(R) &= \sum_{l=1}^5 a'_l R^l
\label{Esigma}
\end{eqnarray}
(the coefficients $a_l$ and $a'_l$ are given
in Table III of Ref. [\onlinecite{said}]).
To make contact with our hamiltonian, we calculate the singlet
triplet splitting applying $H_\pi$ to the ethylene molecule.
Let us denote the singlet and triplet lowest energy states as:
\begin{eqnarray}
\vert s \rangle &=& \cos\theta \frac{
\vert a \overline b\rangle -
\vert \overline a b\rangle}{\sqrt 2} + \sin\theta\frac{
\vert a \overline a\rangle +
\vert \overline b b\rangle}{\sqrt 2}\\
\vert t \rangle &=& \frac{
\vert a \overline b\rangle +
\vert \overline a b\rangle}{\sqrt 2}
\end{eqnarray}
It is easy to show that $^3E=\langle t\vert H_\pi \vert t \rangle = 0 $
while the singlet energy is:
\begin{eqnarray}
&^1E(R) =  \langle s\vert H_\pi \vert s \rangle
= \frac{1}{2}\left[(\gamma_0-\gamma_{1,2}(R))\right.
\cr\cr
&\left.
-\sqrt{\left(
\gamma_0-\gamma_{1,2}(R)\right)^2 + 16 \beta_{1,2}^2(R)}\right]
\end{eqnarray}
where the $\gamma_{ij}$ are given in Eq. 2.4.
This is to be equated to $-2g(R)$ and solved to give the dependence
of $\beta \equiv \beta_{1,2}$ on the bond length:
\begin{equation}
\beta(R) = - \sqrt{g^2(R) + \frac{g(R)}{2} \,\frac{\gamma_0^2 R}
{1 + \gamma_0 R} }
\end{equation}
Using the Mataga Nishimoto prescription for $\gamma_0$, we obtain for example
$\beta =-2.936160 eV$ when $R = 1.40 \AA$.

\begin{figure}
\centerline{
\epsfbox{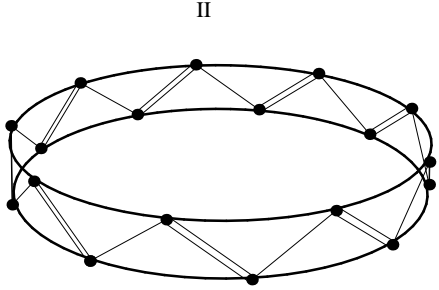}}
\caption{
Structure II: all C atoms are on the surface of a cylinder, 
all bond angles are equal to $120^o$.\label{fig:cylinder}}
\end{figure}

\section{DMRG algorithm and boundary conditions}

We are now ready to implement the DMRG procedure.
At every stage the actual sites are decomposed into sub-blocks
forming a ''superblock'' $A\, \bullet\, \bullet\, B$ (where $\bullet$ 
denotes a single site). Initially $A$ and $B$ are single site blocks.
In order to enlarge the blocks, the reduced density matrix of the 
''system'' $A\, \bullet$ is computed and its most representative states
are retained to
define a basis of the new block $A$ for the next iteration.
The blocks $B$ are obtained from the blocks $A$ through reflection.

Once the total size of the whole system is reached,
the algorithm may be continued with an asymmetric decomposition
$A\, \bullet\, \bullet\, B$
with $A$ and $B$ different from one another.
We can increase $A$ and decrease $B$ and viceversa,
optimizing the states that describe the blocks.
This part is called finite-size algorithm, opposed to the first part
called infinite-size algorithm (see Refs. [\onlinecite{white1,white2,fano}]
for details).

Generally in DMRG calculations open boundary conditions are imposed
because the decay of the density matrix eigenvalues is faster with
respect to the case of periodic boundary conditions and
therefore the convergence is more accurate. We have tested that this
difference is still present  even with a long ranged potential,
as in our case.
When the system is finite the properties of the system depend on the
choice of the boundary conditions.
We put our system on a ring (structure I shown in Fig. 
\ref{fig:circumference}), since it is a
common experience in finite-size calculations that the absence of
boundaries ensures a nice  (smooth) scaling of the energy with 
the system size.  
Let us denote by $r_c$ the circle  radius.
    We want to consider only closed shell systems, since the 
behaviour of the closed shell energies with respect to N
is quite smooth \cite{stef3} 
(the same is also true for low lying excited states
that have an open shell structure and the same symmetry).

To obtain a closed shell state, we
use periodic boundary conditions (PBC)
for $N=4\nu + 2$ and antiperiodic boundary conditions (ABC) for $N=4\nu$
(denoting by $c_i$ the annihilation operator corresponding to the site $i$, 
ABC amount to put $c_N = - c_0$). In fact the allowed pseudomomenta are:
\begin{equation}
 k_n = \left\{ \begin{array}{ll}
\frac{\pi}{N} 2 n & \hbox{ PBC} \\
\frac{\pi}{N} (2 n + 1) &\hbox{ ABC}
\end{array}
\right.
\quad\quad -\frac{N}{2} \leq  n < \frac{N}{2}
\end{equation}

We denote by  $\alpha_{0 1} $ and $\alpha_{1 2} $ the two angles seen from
the center corresponding to the double  and single  bond  respectively,
by $\alpha_{i j} $ the angle between sites $i$ and $j$, by $R_0$ the
average bond-distance and by $\delta$ the dimerization:
\begin{equation}
\delta = \frac { R_{i,i+1} - R_{i-1,i} } {2}  
\end{equation}
From the equations:
\begin{eqnarray}
R_0 - \delta  &=& 2  r_c \sin\left(\frac{\alpha_{0 1}}{2}\right)\\
R_0 + \delta  &=& 2  r_c \sin\left(\frac{\alpha_{1 2}}{2}\right)\\
{ {\alpha_{0 1} + \alpha_{1 2} } \over 2} &=& { {2 \pi }\over N} 
\end{eqnarray}
it is easy to deduce that the circumference radius is given by:
 \begin{equation}
    r_c   =
         \frac{\sqrt{
          R_0 ^2 \cos^2(\frac{\pi}{N}) + \delta^2 \sin^2(\frac{\pi}{N})
        }}
        {\sin(\frac{2\pi}{N})  }
\end{equation}

\begin{figure}
\centerline{
\epsfbox{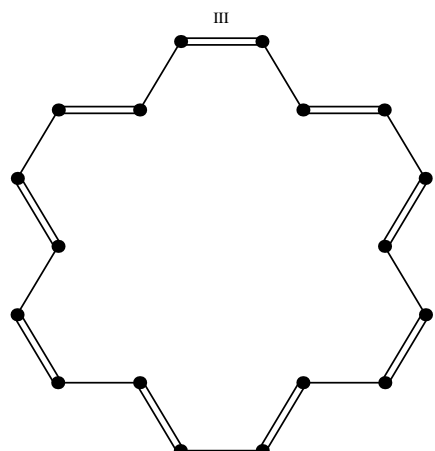}}
\caption{
Structure III: annulene $D_{3h}$ 
geometry, proper of $6(2n+1)$ systems, here shown for N = 18. 
All bond angles are equal to
$120^o$. 
\label{fig:annulene}}
\end{figure}
Other geometries are also possible.
In particular, one can fix the bond angles to  $120^o$.
This value is close to the experimental bond angle of the $sp^2$ hybridized 
carbon in linear polyenes.
Two different geometrical arrangements
of the atoms with bond angles of 
$120^o$ are shown in Fig. \ref{fig:cylinder} and \ref{fig:annulene}.
Imposing periodicity on a linear chain implies 
that atoms are disposed on the surface of a cylinder (structure II shown
in Fig. \ref{fig:cylinder}).
The cylindrical geometry can easily be implemented within
the above DMRG scheme.
The annulene $D_{3h}$ geometry, proper of $6(2n+1)$ systems \cite{Kertesz}
(structure III shown in Fig. \ref{fig:annulene}),
requires a slight modification of the DMRG superblock arrangement, which must
be changed to $\bullet A B \bullet$ due to the lower symmetry 
(translation invariance is lost).
We note that the sigma part of the PPP
hamiltonian does not depend on bond angles.
Therefore the energy difference between the various geometries is purely due to
the distance difference between second and further neighbours in the molecule.

\section{Accuracy test}

There are two ways of improving the accuracy of a DMRG calculation:
increasing the number $m$ of states kept to describe $A$ and performing 
one or more iterations of the finite-size algorithm.
For comparison, we compute two sets of values,  obtained  respectively
with the infinite-size algorithm and $m = 256$ states (DMRG(1)) 
and with the finite-size algorithm
and $m = 512$ states in the last iteration (DMRG(2)).
Even when we limit ourselves to the infinite-size algorithm
we must use from the very
beginning the hamiltonian coefficients that describe the final system,
because the distances depend on the final size.

We test both methods against FCI calculations for $N = 14$.
As Table \ref{table1} shows, 
the absolute error of DMRG(1) is of order $1 \times 10^{-2}$ eV,
while the absolute error for DMRG(2) is of order $2 \times 10^{-4}$ eV.

\begin{table}
\caption{Energy results for N = 14: the $\pi$ energies calculated
with DMRG are compared to  FCI
values. In DMRG(1) we use the infinite-size algorithm and keep 256 states.
In DMRG(2) we use the finite-size algorithm and keep 
512 states in the third (last) iteration.  $R_0$ and $\delta$ are in 
$\AA$, energies in $eV$.
}
\begin{tabular}{ccccc}
$R_0$  & $\delta$ & $E_{\pi}[DMRG(1)]$ & $E_{\pi}[DMRG(2)]$ & $E_{\pi}[FCI]$         \\
\tableline
 1.390 & 0.0000   & -35.678793 & -35.685545 &  -35.685710 \\
       & 0.0100   & -35.718946 & -35.725648 &  -35.725808 \\
       & 0.0200   & -35.835913 & -35.842554 &  -35.842695 \\
 1.400 & 0.0000   & -34.873826 & -34.880260 &  -34.880426 \\
       & 0.0100   & -34.913695 & -34.920161 &  -34.920230 \\
       & 0.0200   & -35.029901 & -35.036097 &  -35.036154 \\
 1.410 & 0.0000   & -34.083595 & -34.089740 &  -34.089900 \\
       & 0.0100   & -34.123179 & -34.129256 &  -34.129409 \\
       & 0.0200   & -34.238365 & -34.244233 &  -34.244367 \\
\tableline
\end{tabular}
\label{table1}
\end{table}

Nonetheless, dimerization and average bond length
are in substantial agreement:
a polynomial fit of the data (even in $\delta$) gives
$R_0=1.40414\AA$ for DMRG(1) ,
$R_0=1.40410\AA$ for DMRG(2) and FCI, and $\delta$ vanishing in all 
three cases.
Anyway we do not learn much from the $N=14$ case 
since the difficult task 
is the determination of $\delta$.
We note that the total energy is not
very sensitive with respect to variations of $\delta$.

For $N=16$, $\delta$ is non-vanishing. In this case
we test the accuracy of the method by comparing energies 
obtained  exchanging double and single bonds.
The hamiltonian with (anti)periodic boundary conditions
is invariant with respect to the exchange of double and single bonds.
Our  arrangement used in DMRG is not invariant:
the bonds connecting the blocks are always asymmetric with respect to exchange
of single and double bonds
and also the bonds internal to blocks $A$ and $B$ 
are asymmetric for certain numbers of sites.
So the symmetry will only be recovered
if convergence is accurately achieved. 
Let $E_{\pi}(+\delta)$ 
($E_{\pi}(-\delta)$) denote the $\pi$-energy obtained with 
$R_{0,1}$  a double (single) bond.
Table \ref{table2} shows these sets of energies for the case  $N=16$.

We find that DMRG(1) does not provide
a very  accurate  determination of $\delta$:
minimizing over the set of energies $E(+\delta)$ we obtain the values
$R_0  = 1.40593\AA$, $\delta = 0.0236\AA$, while
minimizing over the set of energies $E(-\delta)$ we 
get a quite different minimum
$R_0 = 1.40599\AA$,  $\delta = 0.0240\AA$.
 DMRG(2) improves the 
result considerably:
from the set of energies $E(+\delta)$ we get
$R_0 = 1.40577\AA$, $\delta = 0.0227(6)\AA$,
from $E(-\delta)$ we get
$R_0 = 1.40577\AA$, $\delta = 0.0227(2)\AA$.
Thus we see that the finite-size algorithm proves to be very useful.
In order to spare computing time in this case, it is worth
implementing a transformation of the wavefunction 
that gives a guess for starting the next iteration, as 
suggested by White\cite{white-wf}.

\begin{table}
\caption{Energy results for N = 16: 
comparison of DMRG(1) and DMRG(2) $\pi$-energies
$E_{\pi}(+\delta)$ ($E_{\pi}(-\delta)$) 
obtained by choosing
$R_{0,1}$  a double (single) bond respectively.
$E_{\pi}(+\delta)$  and $E_{\pi}(-\delta)$ 
are very close for the case DMRG(2).
$R_0$ and $\delta$ are in 
$\AA$, energies in $eV$. 
}
\begin{tabular}{cccccc}
\multicolumn{2}{c}{}&\multicolumn{2}{c}{DMRG(1)}&
\multicolumn{2}{c}{DMRG(2)}\\
\hline
$R_0$ & $\delta$ &  $E_{\pi}(-\delta)$ & $E_{\pi}(+\delta)$ & $E_{\pi}(-\delta)$  &$E_{\pi}(+\delta)$ \\
\hline
1.400 &   0.020  & -39.908840 & -39.908750 &   -39.920105  & -39.920102\\
      &   0.025  & -40.014971 & -40.014343 &   -40.024754  & -40.024635\\
      &   0.030  & -40.138842 & -40.138400 &   -40.147548  & -40.147519\\
1.405 &   0.020  & -39.456517 & -39.456962 &   -39.467522  & -39.467413\\
      &   0.025  & -39.562115 & -39.561796 &   -39.571654  & -39.571722\\
      &   0.030  & -39.685307 & -39.685028 &   -39.693748  & -39.693736\\
1.410 &   0.020  & -39.008484 & -39.008540 &   -39.019108  & -39.019058\\
      &   0.025  & -39.113436 & -39.112946 &   -39.122670  & -39.122689\\
      &   0.030  & -39.235946 & -39.235557 &   -39.244184  & -39.244095\\
\hline
\end{tabular}
\label{table2}
\end{table}

\begin{figure}
\centerline{\epsfbox{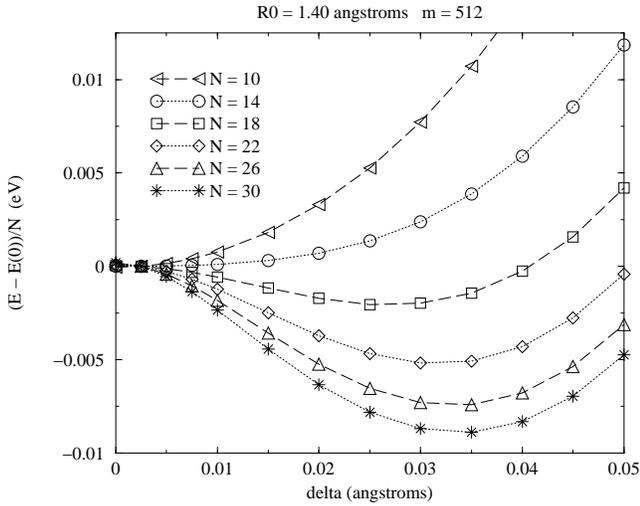}}
\caption{
Total energies at fixed $R_0 = 1.4\AA$. 
Results are obtained with 
DMRG(2). $\delta$ vanishes for $N \leq 14$.
\label{fig:multimin}}
\end{figure}

\section{Results and discussion}

In Fig. \ref{fig:multimin}
we show the behaviour of the energy versus $\delta$
for fixed $R_0 = 1.40\AA$, using DMRG(2). 
In contrast with the case of open boundary conditions used in Ref.
[\onlinecite{lepetit}],  we find $\frac{\partial E}{\partial \delta}\vert
_{\delta = 0} \approx 0$, as expected.
We observe the transition from a non-dimerized minimum for $N \leq 14$
to a dimerized minimum for $N \geq 18$.

\begin{figure}
\centerline{\epsfbox{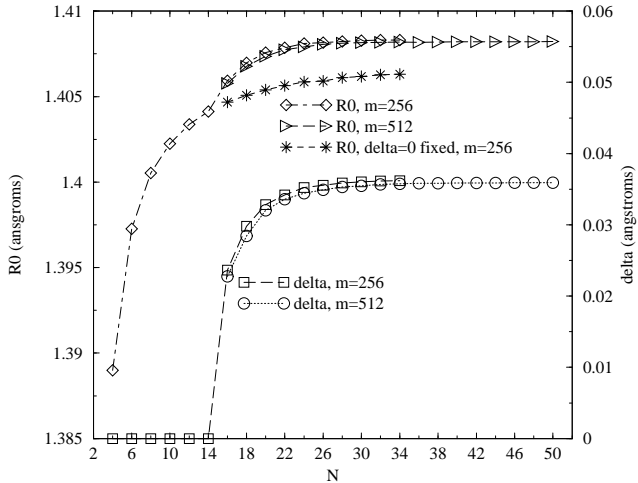}}
\caption{
Average bond
length and dimerization, as determined from minimization 
of total energies with DMRG(1) (m=256 states) and DMRG(2) (m=512 states).
Data denoted with a star indicate values of the average bond length
obtained from minimization with the constraint $\delta=0$. 
\label{fig:rd_min}}
\end{figure}

\begin{figure}
\centerline{\epsfbox{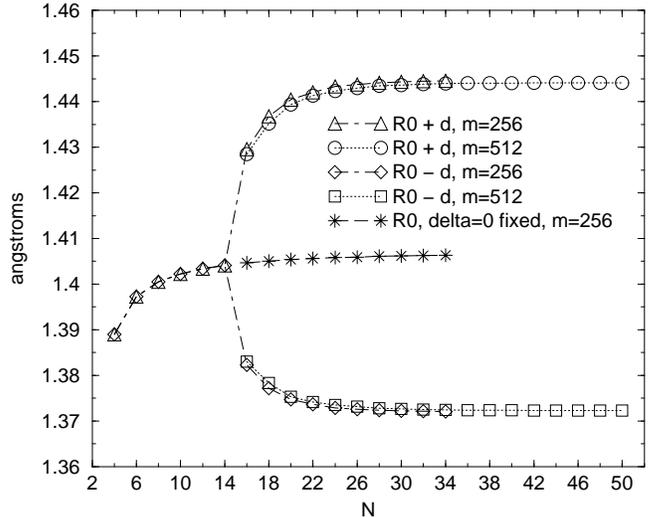}}
\caption{
Single and double bond lengths 
as obtained from energy minimization with both DMRG(1) and DMRG(2).
\label{fig:forcone}}
\end{figure}

We perform two-dimensional minimization (energy versus $\delta$ and $R_0$)
with both DMRG(1) and DMRG(2) as shown in Fig. \ref{fig:rd_min}, 
and \ref{fig:forcone}.
We find  a discontinuity in the slope of the curve of $R_0$ versus
$N$ in correspondence of the transition
from a non-dimerized minimum to a dimerized one ($N > 14$).

In fact it can be seen from Fig. \ref{fig:rd_min} that the curve 
of $R_0$ values obtained fixing $\delta = 0$ shows no discontinuity.
The asymptotic values for $R_0$ and $\delta$ are approximately
given by $1.408(3)\AA$ and $0.036(0)\AA$.
The critical value $N = 14$ was also found by Paldus and coworkers in 
his VB calculation\cite{Paldus1} as well as in the previous studies of
the series on dimerization of PPP cyclic polyenes\cite{Paldus}.
The hamiltonian used has the same two-body part, but we use 
a different prescription for the dependence of the hopping integral $\beta$
and the sigma energy $E_{\sigma}(R_{ij})$ upon the interatomic distance.
Recent  DFT calculations\cite{Kertesz} on the annulenes
$C_{N}H_{N}$, with $N \leq 66$  give a critical value $N = 30$, confirming
and extending previous results of MP2\cite{Jiao} and DFT 
calculations\cite{Kertesz1}.
A direct comparison of the results is rather problematic due to 
the difference between the models. In our case we have a simplified
hamiltonian treated at a high level of approximation by the DMRG
method. On the other hand, the DFT calculation relies
upon a more realistic all electron \emph{ab-initio} hamiltonian,
with a poorer treatment of the long ranged electron correlations.

\begin{figure}
\centerline{\epsfbox{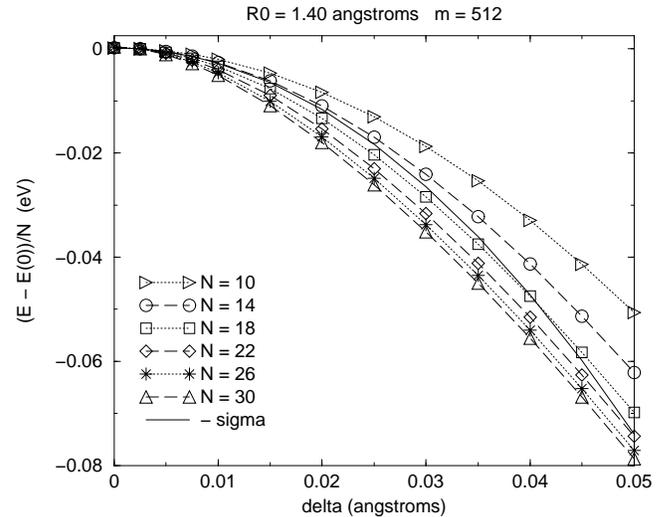}}
\caption{
Comparison of energies versus $\delta$
at fixed $R_0 = 1.4\AA$.
The symbols denote $\pi$ energies per carbon atom
for $N$ between 10 and 30,
the solid line denotes the opposite of the $\sigma$ 
energy per carbon atom.
Results are obtained with DMRG(2).\label{fig:esigma_pi}}
\end{figure}

The location of the minimum of the energy as a function
of $R_0$ and $\delta$ is the result of a  balance of 
the $E_{\sigma}$ and
$E_{\pi}$ energies, as can be seen from Fig. \ref{fig:esigma_pi},
and one might wonder whether a small variation of one of
the two opposite terms could significantly change the results.
The transition from the non alternating to 
the dimerized regime can be followed
by considering the second derivative of the total energy 
(per carbon atom) with respect to $\delta$,
$E"(\delta)=E"_{\sigma}(\delta)+E"_{\pi}(\delta)$.
In the non alternating situation we have a minimum 
at $\delta=0$, i.e. $E"(0)>0$, while in the
dimerized one we have two symmetrical minima at 
$\delta \neq 0$, i.e. $\delta=0$ is now
a local maximum and $E"(0)<0$.
Therefore we can study the sign of the second derivative 
of the energy per carbon atom as a function of
the number of atoms $N$:  at the transition point 
$E"$ must be vanishing. In Table \ref{table_deriv} we
report  the values of 
$E"_{\pi}$ per carbon atom for $N=10, 14,  \ldots 30$ 
together with $E"_{\sigma}$ per carbon atom
(independent from $N$); it is easily seen that the transition 
occurs between $N=14$ and $N=18$.
From these data it can be estimated that in order 
to have the transition at $N=30$, we 
need to alter the ratio $E"_{\sigma} / E"_{\pi}$ by a
factor $ \approx 2$, which cannot
be considered a small adjustment of the model.

In order to check the influence of geometry on these results, we
choose the special case $N=18$, and we compare energies
obtained with the three different geometries of Fig. 
\ref{fig:circumference}- \ref{fig:annulene}. 
The energies  are shown in  Table \ref{table3}.
The  geometry does not affect the dimerization in a sensible way,
although the value of $\delta$ is slightly reduced by going from 
({\bf I}) to ({\bf III}). The values are the following:
\begin{tabbing}
(III) annulene: \= $R_0 = 1.4063(1) \AA$,\= $\quad\delta = 0.0271(6) \AA$\kill
(I) ring: \> $R_0 = 1.4067(6) \AA$,\> $\quad\delta = 0.0290(0) \AA$\\
(II) cylinder: \> $R_0 = 1.4064(7) \AA$,\> $\quad\delta = 0.0276(1) \AA$\\
(III) annulene: \> $R_0 = 1.4063(1) \AA$,\> $\quad\delta = 0.0271(6) \AA$\\
\end{tabbing}
Table \ref{table3} shows that 
the annulene geometry is variationally preferred. 
However, the $\sigma$ energy is completely 
unaffected by these geometry changes.
The small differences in the values of $R_0$ and $\delta$ are only due to
the $\pi$ part of the Hamiltonian.

\begin{table}
\caption{
Second derivatives of $\sigma$ and $\pi$ energies per carbon atom 
at zero dimerization,
as obtained from DMRG(2) data. $E''_{\sigma}$ does not depend on $N$.
Derivatives  are in atomic units.
}
\begin{tabular}{ccc}
$N$ & $E''_{\sigma}(\delta = 0)$ & $ E''_{\pi}(\delta = 0)$ \\
\tableline
10  & 0.6082   & -0.44(0) \\
14  &   ``     & -0.59(1) \\
18  &   ``     & -0.74(9) \\
22  &   ``     & -0.91(1) \\
26  &   ``     & -1.06(7) \\
30  &   ``     & -1.21(2) \\
\tableline
\end{tabular}
\label{table_deriv}
\end{table}

\begin{table}
\caption{Energy results for N = 18: 
comparison of $\pi$-energies from DMRG(2) with different geometries,
shown in Fig. 1, 2  and 3. $R_0$ and $\delta$ are in 
$\AA$, energies in $eV$.
}
\begin{tabular}{ccccc}
\hline
$R_0$ & $\delta$ &  ring (I) & cylinder (II) & annulene (III) \\
\hline
1.400 &   0.020  & -44.826681 & -44.913191 & -44.951474 \\
      &   0.025  & -44.952568 & -45.037015 & -45.074791 \\
      &   0.030  & -45.098006 & -45.180696 & -45.217941 \\
1.405 &   0.020  & -44.318921 & -44.404909 & -44.442656 \\
      &   0.025  & -44.444324 & -44.528098 & -44.565347 \\
      &   0.030  & -44.589043 & -44.670964 & -44.707697 \\
1.410 &   0.020  & -43.815950 & -43.901321 & -43.938530 \\
      &   0.025  & -43.940718 & -44.023869 & -44.060595 \\
      &   0.030  & -44.084491 & -44.165918 & -44.202143 \\
\hline
\end{tabular}
\label{table3}
\end{table}

\end{document}